\documentclass[]{elsart}
\usepackage{fullpage}

 \usepackage{graphicx}

\usepackage{amssymb}

\begin{document}

\begin{frontmatter}



\title{Quantum manipulation and simulation using Josephson junction arrays}
\author{Xingxiang Zhou}
\ead{xizhou@yahoo.com}
\author{and}
\author{Ari Mizel}
\ead{ari@phys.psu.edu}
\address{Department of Physics, The Pennsylvania State University,
University Park, PA 16802, USA}




\begin{abstract}
We discuss the prospect of using quantum properties of large scale 
Josephson junction arrays for quantum manipulation and simulation. 
We study the collective vibrational quantum modes of a Josephson junction 
array and show that they provide a natural and practical method for 
realizing a
high quality cavity for superconducting qubit based QED. We 
further demonstrate that by using Josephson junction arrays we can simulate
a family of problems concerning spinless electron-phonon and 
electron-electron interactions.
These protocols require no or few controls over the Josephson junction
array and are thus relatively easy to realize given currently available
technology.
\end{abstract}

\end{frontmatter}



Superconducting device based solid state qubits \cite{Makhlin01} 
are attractive because of their inherent scalability. Microwave 
spectroscopy and long lived population oscillation 
consistent with single \cite{Nakamura99,Friedman00,Mooij00,Vion02,Yu02,Martinis02,Mooij03} and two qubit quantum states \cite{Pashkin03,Berkley03,Izmalkov04,Mooij04} 
have been observed experimentally.


Recently, a new approach to scalable superconducting quantum computing
analogous to atomic cavity-QED was studied theoretically 
\cite{Falci03,Blais03,Zhou04,Blais04} and implemented experimentally 
\cite{Wallraff04,Xu04}. This new approach opens the possibility of 
applying methods and principles from the rich field of atomic QED in
solid state quantum information processing.

One practical problem of solid state qubit based QED is the realization 
of a high quality resonator to which many qubits can couple. A lumped
element on-chip LC circuit, such as that used in \cite{Xu04}, 
suffers from dielectric loss of the capacitor and ac
loss of the superconductor \cite{Kadin99}. A high quality resonator 
can be realized with a co-planar waveguide if high quality substrate  
material (such as sapphire) is used to minimize the loss 
\cite{Mazin02,Day03}. 
A high quality, low leakage Josephson junction provides a natural and 
easy realization of a high quality resonator 
due to its high quality tri-layer structure \cite{Rando95}, however only
a few qubits can couple to such a single junction resonator 
\cite{Falci03,Blais03}.

In this work, we study the quantum dynamics of a Josephson junction 
array and show that the ``phonon modes'' corresponding to the small 
collective vibrations of the junction phases can be used to 
realize a high quality resonator to which many superconducting 
qubits can couple. The resultant structure is analogous to the ion
trap quantum computer in which qubits communicate through the phonon 
modes \cite{Cirac95}. We further show that, by using a properly coupled 
superconducting qubit array we can simulate a family of problems 
involving spinless electron-phonon and electron-electron interactions.

\begin{figure}[h]
    \includegraphics[width=3.2in, height=2.4in]{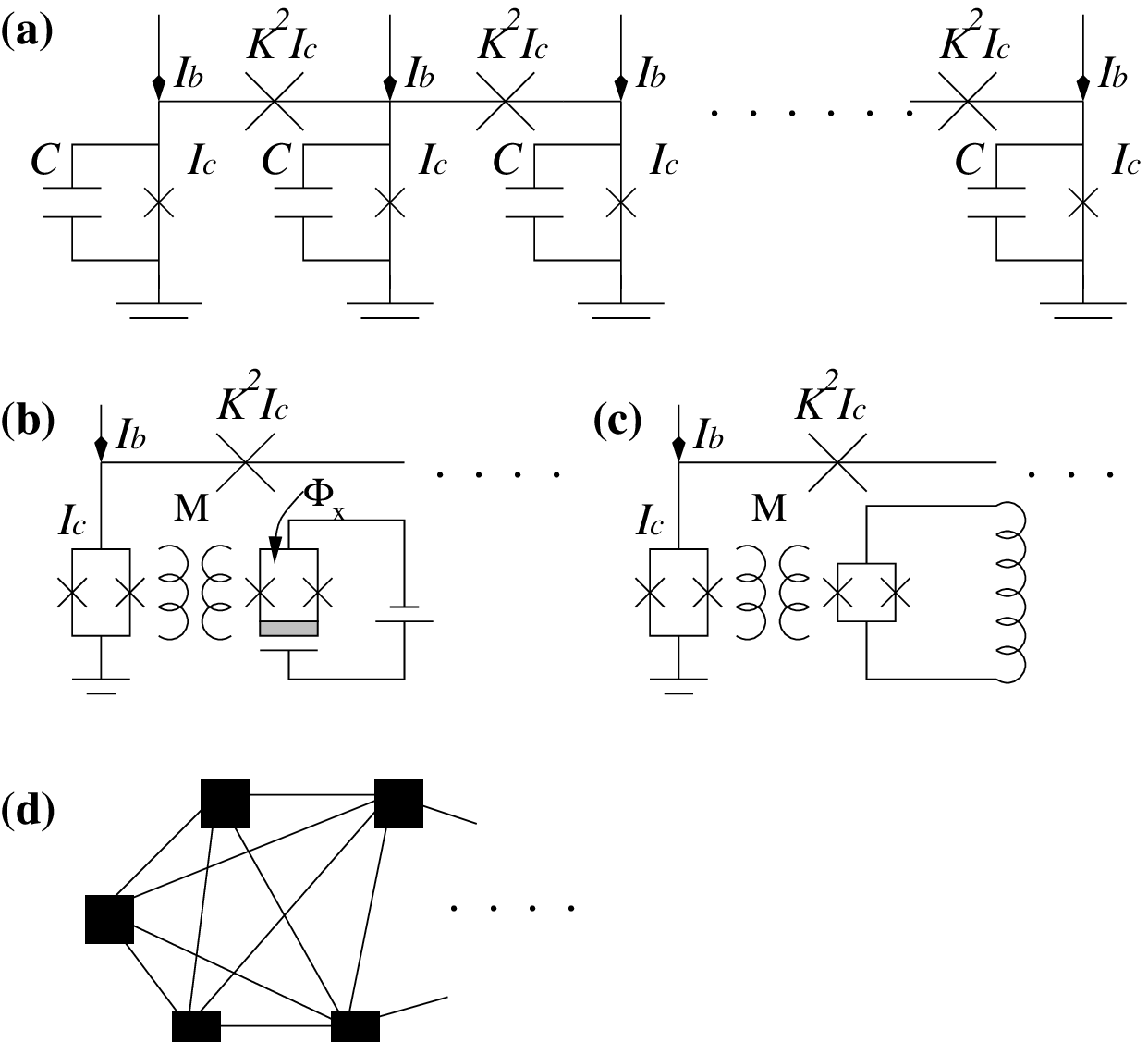}
    \caption{(a) A simple Josephson junction array consisting of $N$ 
      vertical and $N-1$ horizontal junctions. (b) A charge  
      qubit coupled to a vertical junction (realized with a small
      self-inductance
      dc-SQUID) in the Josephson junction array in (a). (c) An rf-SQUID
      qubit coupled to the Josephson junction array. (d) A more 
      complex design than the simple array in (a). Each dot represents 
      a superconducting island which is grounded by a Josephson junction 
      (grounding junctions not shown). Each and every pair of islands are 
      coupled by a Josephson junction as indicated by the edges connecting 
      the dots.
    }
    \label{fig:Array}
\end{figure}

Consider the simple Josephson junction array shown in Fig. 
\ref{fig:Array} (a). Denote the phases across the vertical junctions 
$\theta_0$, $\theta_1$, ..., $\theta_{N-1}$ (the phase of the ground is set
to 0). The capacitance of the vertical junctions is $C$. The horizontal 
junctions are much bigger in size than the vertical ones, and their 
critical current is $K^2$ times that of the vertical junctions ($I_c$), 
where $K\gg 1$. In practice, all junctions can be realized by low 
self-inductance dc-SQUIDs to allow tuning of their critical currents;
therefore $K^2$ is not necessarily equal to the ratio of the junction 
sizes. Each vertical junction is biased by a current $I_b< I_c$ to 
suppress its plasma frequency. The 
geometric inductance is very small and neglected.

The potential energy of the Josephson junction array in Fig. 
\ref{fig:Array} is a sum of the Josephson tunneling energies of the 
junctions, given by
\begin{equation}
V=-E_J\sum_{i=0}^{N-1} (i_b\theta_i + \cos{\theta_i}) 
-K^2E_J\sum_{i=0}^{N-2} \cos{(\theta_i- \theta_{i+1})}, \nonumber
\label{eq:H_CB}
\end{equation}
where $E_J=I_c\Phi_0/2\pi$ and $i_b=I_b/I_c$. The equilibrium values for 
the junction phases are determined by $\partial V/ \partial \theta_i =0$, 
which is just the current conservation condition at each node: $i_b 
-\sin{\theta_i} -\sin{(\theta_i -\theta_{i-1})} -\sin{(\theta_i 
-\theta_{i+1})} =0$. (When $i=0$ or $N-1$ there is only one horizontal 
current.) In our setup, the equilibrium values for the phases are
$\theta_i^{(0)} =\theta^{(0)} =\arcsin{i_b}$ for $i=0,1,...N-1$.
If we are interested in the small oscillations of the phases, we 
follow the standard procedure expanding $V$ to second order, 
$V=\frac{1}{2}\sum_{i,j=0}^{N-1} V_{ij}\varphi_i\varphi_j$, where $\varphi_i
=\theta_i -\theta_i^{(0)}$ is the small displacement of the phase from 
its equilibrium value and
\begin{equation}
\frac{(V)_{ij}}{E_JK^2}= \delta_{ij} (2+\cos{\theta^{(0)}}/K^2 
-\delta_{i,0} -\delta_{i,N-1}) -\delta_{i,j-1} -\delta_{i,j+1}. \nonumber
\end{equation}
Notice that the horizontal junctions are much larger than the vertical 
ones. In the ``wash board'' analogy this corresponds to a large ``mass''
for the horizontal junctions.
Therefore in calculating the kinetic energy of the system we need only 
consider the vertical junctions (as verified by numerical calculations):
$T=(1/2)C(\Phi_0/2\pi)^2 \sum_{i=0} ^{N-1} \dot\varphi_i^2$. From the 
potential and kinetic energies we can solve for the normal modes of the 
system, whose spectrum is given by $\nu_s= \omega_p\sqrt{1+ (4K^2/
\cos{\theta^{(0)}}) \sin ^2 (s\pi/2N)}$, where the plasma frequency
$\omega_p= \sqrt{2\pi I_c/\Phi_0 C}(1-i_b^2)^{1/4}$ and $s=0, 1, ..., N-1$.
Let the orthonormalized normal mode eigenvectors be denoted $b_i^s$.
The lowest mode, whose frequency is
$\omega_p$, corresponds to the center of mass motion of the phases: 
$b^0 =(1,1,...,1)/\sqrt{N}$. The quantum mechanical properties of the
small collective vibrational modes of the phases can now be evaluated by
introducing the operator $\hat{\varphi_i}= (2\pi/\Phi_0) \sum_{s=0}
^{N-1} \sqrt{\hbar/2C\nu_s} b_i^s (a_s +a_s^{\dagger})$, where $a_s$ 
is the annihilation operator for the $s$th mode.

We propose to use the center of mass motion mode of the Josephson
junction array to couple superconducting qubits, in analogy to ion 
trap quantum computer \cite{Cirac95}. 
Since the center of mass motion mode is an equal 
weight superposition of the junction phases, its quality factor is as
high as that of the individual junctions. 
Therefore, this approach allows us to take advantage of the high 
quality Josephson junctions required for the superconducting qubits to 
realize a high quality resonator. Note in the above we have assumed 
that all Josephson junctions in the array are identical. In reality, 
this will not be the case due to unavoidable fabrication errors. However 
the critical currents of the junctions can be tuned by a magnetic field 
so that the effective Josephson energies of the junctions can be equalized. 
The effect of any residual asymmetry can be estimated by 
perturbation theory. As long as the amplitude of the transition matrix 
element due to the asymmetry is much smaller than the energy gap between 
the center of mass motion mode and higher modes, the energy and 
wavefunction of the center of mass motion mode remain close to  
unperturbed.


In order to implement protocols developed in superconducting qubit based 
cavity-QED \cite{Falci03,Blais03,Zhou04,Blais04}, we need to couple 
the superconducting qubits to the center of mass motion mode of the junction
array in a way such that they can exchange energy. This is shown in Fig. 
\ref{fig:Array} (b) and (c) for both charge and flux qubits. Here each vertical
junction in the junction array is replaced by a small self-inductance
dc-SQUID and coupled inductively to a superconducting qubit. Consider the 
charge qubit whose Hamiltonian is $H_Q =-B^z\sigma^z/2 -B^x\sigma^x/2$, 
where $B^z$ and $B^x$ are determined by the gate voltage and Josephson 
energy of the 
charge qubit. When its energy $B^z$ is tuned close to $\nu_0 =\omega_p$ 
and its dc-SQUID is biased at $\Phi_0/2$ (including the flux due to the 
junction array's bias current $I_b$), the inductive coupling 
results in a coupling Hamiltonian $H_c= -g(a\sigma^+ +a^{\dagger}\sigma^-)$, 
where $g= (M/2)(I_c\cos{\theta^{(0)}})I_c^Q (2\pi/\Phi_0) 
\sqrt{\hbar/2C\omega_p N}$, $M$ is 
the mutual inductance, 
$I_c^Q$ is the critical current of the dc-SQUID junctions of the charge 
qubit, 
and $a$ is the annihilation operator for the center of mass motion mode of the 
junction array. 
In deriving $H_c$, we have used the rotating wave 
approximation to drop terms that oscillate at high frequencies. For 
the flux qubit case shown in Fig. \ref{fig:Array} (b),
we can derive the same
coupling Hamiltonian, with a coupling coefficient which is proportional 
to the mutual inductance and can be evaluated in terms of the qubit 
parameters \cite{Orlando99}. With the above design we then have a 
structure in close analogy to the ion trap quantum computer in which 
the qubits communicate through the center of mass phonon mode. To 
realize a universal quantum computer, we can use either the resonant
\cite{Falci03,Blais03} or dispersive \cite{Falci03,Zhou04,Blais04} interaction 
between the qubits and the junction array mode. The interaction is
switched on and off by tuning the energy of the qubit into and out of
resonance with the junction array.


Typical values for $\omega_p$ and qubit energies can be chosen to be
up to 10GHz.
The coupling strength $g$ can be tens of megahertz \cite{Zhou04,You}. 
When the system scales up ($N$ increases), the energy gap between 
the center of mass motion mode and the higher modes should remain much 
greater than the coupling strength $g$ to avoid excitation of upper 
modes. When $N$ is large, the energy difference between the lowest two 
modes is $\Delta\nu_{01} =(\pi^2K^2/2N^2\cos{
\theta^{(0)}}) \omega_p$. For $K=20$ \cite{Note} and $i_b=0.97$, 
this implies an upper
limit of a few hundred for $N$. To relax this limit, we can use more 
complicated designs than the simple 1d array in Fig. \ref{fig:Array} (a). 
For instance, we can consider a network
of $N$ superconducting islands in which each pair of nodes is coupled by a 
Josephson junction, as shown in Fig. \ref{fig:Array} (d).
Each island is still grounded though a junction whose plasma
frequency is $\omega_p$, and the Josephson energy of the coupling 
junctions is $K^2$ that of the grounding junctions.
In this case, the center of mass motion mode remains at $\omega_p$ and
all higher modes are pushed up to a frequency $\omega_p\sqrt{1+NK^2/\cos{
\theta^{(0)}}}$. 
The number of junctions
required is on the order $N^2/2$.

In the above, we take advantage of the macroscopic quantum behavior of 
the Josephson junction phases to construct an analogy of the ion trap
quantum computer.
We can push this concept further and consider using 
Josephson junction arrays to simulate the dynamics of 
physical systems. 
%
Consider 
a 1d array of qubits. In the spin 1/2 representation, 
the dynamics of the qubits are described by the spin operators $S_i^x$, 
$S_i^y$ and $S_i^z$ whose commutation relations are defined by 
$[S_i^\alpha, S_j^\beta] =(i/2)\delta_{ij}\epsilon_{\alpha\beta\gamma}S_i^\gamma$ 
for $\alpha, \beta, \gamma = x, y, z$ and 
$i, j =0, 1, ...,N-1$. By using a Jordan-Wigner 
transformation \cite{Jordan28}, we can map this qubit array to a 
collection of spinless fermions
annihilated by the operators 
\begin{equation}
f_n =S_n^- K(n), \label{eq:JW}
\end{equation}
where $S_n^{\pm} =S_n^x\pm iS_n^y$ and $K(n)$ is a nonlocal 
operator defined by $K(n) =exp[i\pi\sum_{m=0}^{n-1} f_m^+f_m]
=exp[i\pi\sum_{m=0}^{n-1} (S_m^z+1/2)]$. It can be verified that
$f_n$'s satisfy the anti-commutation relations required for fermion 
operators. The fermion number operator is simply $f_n^{\dagger}f_n =S_n^z 
+1/2$. Now, if we couple the qubits by nearest neighbor XXZ interactions,
the Hamiltonian of the system is $H 
=-J_{xy}\sum_i (S_i^xS_{i+1}^x +S_i^yS_{i+1}^y) +J_z \sum_i
S_i^zS_{i+1}^z$. Under the Jordan-Wigner transformation, the system
Hamiltonian is transformed to
\begin{equation}
H =-(J_{xy}/2)\sum_n (f_n^{\dagger}f_{n+1} +f_{n+1}^{\dagger}f_n) 
+J_z \sum_n
(f_n^{\dagger}f_n- 1/2) (f_{n+1}^{\dagger}f_{n+1}- 1/2).
\end{equation}
As can be seen the coupling in the XY directions
transforms into hoping of the fermions 
between neighboring sites and the Ising coupling causes nearest 
neighbor interactions between the fermions.

\begin{figure}[h]
    \includegraphics[width=3.2in, height=1in]{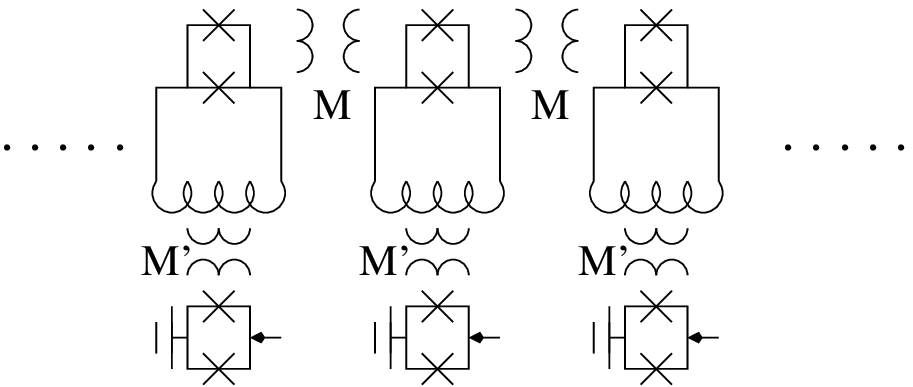}
    \caption{A 1d rf-SQUID qubit array inductively coupled to local
      Josephson junction oscillators. The dc-SQUIDs of 
      the qubits are inductively coupled too. Both $M$ and $M'$ are 
      tunable couplings \cite{Orlando99,Mooij99,Filippov03}. The qubits are
      arranged in a ring structure when periodic boundary conditions
      are required.
    }
    \label{fig:QubitArray}
\end{figure}

An interesting scenario arises when we couple a superconducting 
qubit array and ``phonon modes'' realized by small oscillations of
Josephson junction phases, as shown in Fig. \ref{fig:QubitArray}. 
Since the qubit array can be transformed into a fermion system, 
this allows us to simulate a 
family of solid state problems concerning spinless electron-phonon 
and electron-electron interactions. The macroscopic nature of the
system and the fine control and measurement techniques developed in
superconducting quantum information processing allow us to examine
the fermion dynamics directly and closely to obtain vital 
information about the system. For this purpose, we need to be careful with
a few issues. The first is that in a physical system the 
electron number is conserved; however a $B^x\sigma^x$ term of the 
qubit can flip its basis states defined by the eigenstates of $\sigma^z$. 
According to the Jordan-Wigner transformation Eq.
(\ref{eq:JW}) the flipping of the qubit state corresponds to the
creation or annihilation of a fermion. Therefore, we must keep
$B^x$ of the qubits zero. 
This is realized by using small self-inductance dc-SQUIDs for the qubits 
and biasing them appropriately 
so that the $B^x$ field of the qubit is 
much smaller \cite{Zhou04} 
than other energy 
scales of the system and thus can be neglected for the time scale of 
interest. To further suppress fermion creation and annihilation, we 
may choose $B^z$ to be large, which makes changes in the total number 
of fermions energetically forbidden.
Another point is that 
the coupling between the qubit and the Josephson phase should 
not allow them to exchange excitations, 
since in a physical system no process happens in which
an electron is created (annihilated) and a phonon is annihilated 
(created). 

In the following we focus on 
the 1d Holstein model of spinless electrons, a problem of 
both theoretical and practical interest \cite{McKenzie96,Bursill98}.
The Hamiltonian of the system is
\begin{equation}
H= -t\sum_i (c_i^{\dagger}c_{i+1} +c_{i+1}^{\dagger}c_i) +\omega
\sum_i a_i^{\dagger}a_i -g\sum_i
(c_i^{\dagger}c_i -1/2)(a_i+a_i^{\dagger}),
\label{eq:Holstein}
\end{equation}
where $c_i$ destroys a fermion at site $i$ and $a_i$ destroys a local
phonon of frequency $\omega$. This system at half filling has two 
phases depending on the ratio $g/\omega$. When $g< g_c$, the ground 
state has a metallic (Luttinger liquid) phase. When $g$ exceeds $g_c$, 
it exhibits an insulating phase with charge density wave (CDW) long 
range order \cite{McKenzie96,Bursill98}. Though it can be solved 
analytically in certain limits, for generic parameters the accurate
determination of the critical coupling strength and energy gaps is 
difficult and only limited size systems at half filling have been 
studied numerically \cite{McKenzie96,Bursill98}.

We can construct an artificial realization of the 1d Holstein model using 
a superconducting qubit array and local oscillator modes realized by 
Josephson junctions, 
as shown in Fig. \ref{fig:QubitArray}. For clarity, rf-SQUID 
qubits are shown, whose dc-SQUIDs are coupled by a tunable mutual 
inductance. 
The tunable mutual inductances can be realized by using
a flux transformer interrupted by a dc-SQUID whose critical current 
can be tuned by a flux bias, as discussed in 
\cite{Orlando99,Mooij99,Filippov03}.
As a result of this inter-qubit coupling and the qubit bias, the qubit
array Hamiltonian is $H_Q =-\sum_i B^zS_i^z -J\sum_i S_i^xS_{i+1}^x$,
where $B^z$ is determined by the flux bias of the 
qubits and $J$ is proportional to the tunable mutual inductance $M$. The
local phonon modes are realized by Josephson junctions whose oscillation
frequency is the plasma frequency $\omega_p$. These junctions should 
have a deep potential well in order to guarantee that their behavior is 
close to harmonic even when the excitation number is not small. 
Alternatively, if we are also interested in studying the effect of 
nonharmonicity of the phonon modes on the system behavior 
\cite{Mahan96}, we may use an inductor ($L$) shunted dc-SQUID.
In this case, under the condition 
$LI_c \ll \Phi_0/2\pi$, the Hamiltonian for the phonon mode is 
$Q^2/2C_J +\Phi^2/2L' -(I_c/24)(2\pi/\Phi_0)^3\Phi^4$, where $Q$ and
$\Phi$ are the charge and flux
across the dc-SQUID, $C_J$ is the total junction capacitance 
and the
effective inductance $L'= L(1-2\pi LI_c/\Phi_0)$. 
The anharmonicity can be adjusted by tuning the critical current $I_c$ 
of the dc-SQUID. 

The local Josephson junction oscillators are coupled to the main loops of
the rf-SQUID qubits using a tunable mutual inductance as shown in Fig.
\ref{fig:QubitArray}. This introduces a coupling Hamiltonian proportional 
to $S^z\Phi$, with a coupling coefficient proportional to the mutual 
inductance $M'$\cite{Orlando99}. After
quantizing the local phonon modes and applying the Jordan-Wigner 
transformation, 
we obtain the Hamiltonian of the 1d Holstein model Eq. (\ref{eq:Holstein}), 
with the hopping strength $t=J/4$ and the phonon frequency $\omega 
=\omega_p$. 
Since the qubit-qubit and 
qubit-phonon interactions are realized with tunable couplings, 
the ratio of $t$ and $g$ to $\omega$ in the Holstein model can be tuned 
in a wide range. This allows us to explore a large phase space of the 
system. 
To simulate the 1d Holstein model, we first set the number of 
spinless fermions in the system. This is done by initializing half the 
qubits in the $S^z=1/2$ state and the other half in the $S^z=-1/2$ 
state (assuming we are interested in the half filling case), which can 
be accomplished by applying strong $B^z$ fields to the qubits or by 
performing measurements. 
Then $t$ and $g$ are turned on (by tuning the inter-qubit and qubit phonon
couplings) slowly until the intended point $(t,g,\omega)$
in the phase space is reached. 
The phase of the system can be inferred by measuring the fermion numbers
$ f_n^\dagger f_n$, or $S_n^z$, which are independent of $n$ (1/2 for half 
filling) for the metallic phase and varying with $n$ for the insulating phase.
The excitation energy of the system can be determined by spectroscopic 
methods like in \cite{Xu04}.



In conclusion, we have shown that properly engineered Josephson junction
arrays can be used for quantum manipulation and simulation. The discussed
protocols require only limited control of the system and provide 
interesting opportunities for superconducting quantum information 
processing.

The authors acknowledge financial support from the Packard Foundation.





\begin{thebibliography}{00}

\bibitem{Makhlin01} Y. Makhlin, G. Sch\"{o}n, and A. Shnirman,
Rev. Mod. Phys. {\bf 73}, 357 (2001).
\bibitem{Nakamura99} Y. Nakamura, Y. A. Pashkin, and J. S. Tsai, Nature 
(London) {\bf 398}, 786 (1999).
\bibitem{Friedman00} J. R. Friedman, V. Patel, W. Chen, S. K. Tolpygo, 
and J. E. Lukens, Nature {\bf 406}, 43 (2000).
\bibitem{Mooij00} C. H. van der Wal, A. C. J. ter Haar, 
F. K. Wilhelm, R. N. Schouten, C. J. P. M. Harmans, T. P. Orlando, S. Lloyd, 
and J. E. Mooij, Science {\bf 290}, 773 (2000).
\bibitem{Vion02} D. Vion, A. Aassime, A. Cottet, P. Joyez, H. Pothier, 
C. Urbina, D. Esteve, and M. H. Devoret, Science {\bf 296}, 886 (2002).
\bibitem{Yu02} Y. Yu, S. Han, X. Chu, S-I.Chu, and 
Z. Wang, Science {\bf 296}, 889 (2002).
\bibitem{Martinis02} J. M. Martinis  S. Nam, J. Aumentado, and 
C. Urbina, Phys. Rev. Lett. {\bf 89}, 117901 (2002).
\bibitem{Mooij03} I. Chiorescu, Y. Nakamura, C. J. P. M. Harmans, 
and J. E. Mooij, Science {\bf 299}, 1869 (2003).
\bibitem{Pashkin03} Y. A. Pashkin, T. Yamamoto, O. Astafiev, Y. Nakamura, 
D. V. Averin, and J. S. Tsai, Nature (London) {\bf 421},
823 (2003); T. Yamamoto, Yu. A. Pashkin, O. Astafiev, Y. Nakamura, and 
J. S. Tsai, {\it ibid.}, {\bf 425}, 941 (2003).
\bibitem{Berkley03} A. J. Berkley, H. Xu, R. C. Ramos, M. A. Gubrud, 
F. W. Strauch, P. R. Johnson, J. R. Anderson, A. J. Dragt, C. J. Lobb, 
and F. C. Wellstood, Science {\bf 300}, 1548 (2003).
\bibitem{Izmalkov04} A. Izmalkov, M. Grajcar, E. Il'ichev, Th. Wagner, 
H.-G. Meyer, A. Yu. Smirnov, M. H. S. Amin, A. M. van den Brink, and 
A. M. Zagoskin, Phys. Rev. Lett. {\bf 93}, 037003 (2004).
\bibitem{Mooij04} I. Chiorescu, P. Bertet, K. Semba, Y. Nakamura, 
C. J. P. M. Harmans, and J. E. Mooij, Nature (London) {\bf 431}, 
159 (2004).


\bibitem{Falci03}F. Plastina and G. Falci, Phys. Rev. B {\bf 67}, 
224514 (2003).
\bibitem{Blais03}A. Blais, A. M. van den Brink, and A. M. Zagoskin, 
Phys. Rev. Lett. {\bf 90}, 127901 (2003).
\bibitem{Zhou04} X. Zhou, M. Wulf, Z-W. Zhou, G-C. Guo, and M. J. 
Feldman, Phys. Rev. A {\bf 69}, 030301R (2004).
\bibitem{Blais04} A. Blais, R-S. Huang, A. Wallraff, S. M. Girvin, 
and R. J. Schoelkopf, Phys. Rev. A {\bf 69}, 062320 (2004).
\bibitem{Wallraff04} A. Wallraff, D. I. Schuster, A. Blais, L. Frunzio, 
R-S. Huang, J. Majer, S. Kumar, S. M. Girvin, and R. J. Schoelkopf, 
Nature (London) {\bf 431}, 162 (2004).
\bibitem{Xu04} H. Xu, F. W. Strauch, S. Dutta, P. R. Johnson, 
R. C. Ramos, A. J. Berkley, H. Paik, J. R. Anderson, A. J. Dragt, 
C. J. Lobb, and F. C. Wellstood, Phys. Rev. Lett. {\bf 94}, 027003 (2005).
\bibitem{Kadin99} A. M. Kadin, ``Introduction to superconducting
  circuits," Wiley-Interscience, 1999.
\bibitem{Mazin02} B. A. Mazin, P. K. Day, H. G. LeDuc, A. Vayonakis, 
and J. Zmuidzinas, Proc. SPIE {\bf 4849}, 283 (2002).
\bibitem{Day03} P. K. Day, H. G. Leduc, B. A. Mazin, A. Vayonakis, 
and J. Zmuidzinas, Nature {\bf 425}, 817 (2003). 


\bibitem{Rando95} N. Rando, P. Videler, A. Peacock, A. van Dordrecht, 
P. Verhoeve, R. Venn, A. C. Wright, and J. Lumley, J. Appl. Phys. {\bf 77},
4099 (1995).
\bibitem{Cirac95} J. I. Cirac and P. Zoller, Phys. Rev. Lett. 
{\bf 74}, 4091 (1995).
\bibitem{Orlando99} T. P. Orlando, J. E. Mooij, L. Tian, C. H. van der 
Wal, L. S. Levitov, S. Lloyd, and J. J. Mazo, Phys. Rev. B. 
{\bf 60}, 15398 (1999).
\bibitem{You} J. Q. You, C-H. Lam, and H. Z. Zheng, Phys. Rev. B {\bf 63}, 
180501R (2001). J. Q. You, J. S. Tsai, and F. Nori, Phys. Rev. Lett. 
{\bf 89}, 197902 (2002).
\bibitem{Note}
Notice as mentioned earlier 
$K^2$ is not necessarily the ratio of junction sizes since magnetic 
suppression of critical currents is usually needed. 

\bibitem{Jordan28} P. Jordan and E. Wigner, Z. Phys. {\bf 47}, 631 (1928).
\bibitem{Mooij99} J. E. Mooij, T. P. Orlando, L. Levitov, L. Tian, 
C. H. van der Wal, and S.Lloyd, Science {\bf 285}, 1036 (1999).
\bibitem{Filippov03} T. V. Filippov, S. K. Tolpygo, J. Mannik, and 
J. E. Lukens, IEEE Trans. Appl. Supercond. {\bf 13}, 1005 (2003).

\bibitem{McKenzie96} R. H. McKenzie, C. J. Hamer, and D. W. Murray, 
Phys. Rev. B. {\bf 53}, 9676 (1996).
\bibitem{Bursill98} R. J. Bursill, R. H. McKenzie, and C. J. Hamer, 
Phys. Rev. Lett. {\bf 80}, 5607 (1998).
\bibitem{Mahan96} J. K. Freericks, M. Jarrell, and G. D. Mahan, 
Phys. Rev. Lett. {\bf 77}, 4588 (1996).












\end{thebibliography}
\end{document}